\def\av#1{\langle#1\rangle}
\newcommand{\omitit}[1]{}
\documentclass[twocolumn,superscriptaddress,preprintnumbers,amsmath,amssymb]{revtex4}
\usepackage[dvips]{graphicx}
\usepackage{latexsym}
\usepackage{natbib}

\begin{document}
\title{Transport and Percolation Theory in Weighted Networks}
\author{Guanliang Li}
\affiliation{Center for Polymer Studies, Boston University, Boston,
  Massachusetts 02215, USA}
\author{Lidia A. Braunstein}
\affiliation{Departamento de F\'{\i}sica, Facultad de Ciencias Exactas y
  Naturales, Universidad Nacional de Mar del Plata, Funes 3350, (7600) Mar del
  Plata, Argentina}
\affiliation{Center for Polymer Studies, Boston University, Boston,
  Massachusetts 02215, USA}
\author{Sergey V. Buldyrev}
\affiliation{Department of Physics, Yeshiva University, 500 West 185th Street, New York, New York 10033, USA}
\affiliation{Center for Polymer Studies, Boston University, Boston,
  Massachusetts 02215, USA}
\author{Shlomo Havlin}
\affiliation{Minerva Center and Department of
  Physics, Bar-Ilang University, 52900 Ramal-Gab, Israel}
\affiliation{Center for Polymer Studies, Boston University, Boston,
  Massachusetts 02215, USA}
\author{H. Eugene Stanley}
\affiliation{Center for Polymer Studies, Boston University, Boston,
  Massachusetts 02215, USA}

\begin{abstract}

We study the distribution $P(\sigma)$ of the equivalent conductance
$\sigma$ for Erd\H{o}s-R\'enyi (ER) and scale-free (SF) weighted
resistor networks with $N$ nodes. Each link has conductance $g\equiv
e^{-ax}$, where $x$ is a random number taken from a uniform distribution
between 0 and 1 and the parameter $a$ represents the strength of the
disorder. We provide an iterative fast algorithm to obtain $P(\sigma)$
and compare it with the traditional algorithm of solving Kirchhoff
equations. We find, both analytically and numerically, that $P(\sigma)$
for ER networks exhibits two regimes. (i) A low conductance regime for
$\sigma < e^{-ap_c}$ where $p_c=1/\av{k}$ is the critical percolation
threshold of the network and $\av{k}$ is average degree of the network.
In this regime $P(\sigma)$ is independent of
$N$ and follows the power law $P(\sigma) \sim \sigma^{-\alpha}$,
where $\alpha=1-\av{k}/a$.
(ii) A high conductance regime for $\sigma >e^{-ap_c}$ in which we find that
$P(\sigma)$ has strong $N$ dependence and scales as $P(\sigma) \sim
f(\sigma,ap_c/N^{1/3})$.  For SF networks with degree distribution
$P(k)\sim k^{-\lambda}$, $k_{min} \le k \le k_{max}$, we find numerically also two regimes, similar
to those found for ER networks.

\end{abstract}

\maketitle

Recently much attention has been focused on complex networks which
characterize many biological, social, and communication systems
\cite{Albert02,Pastor,Dorogovtsev,Cohen}. The networks are represented
by nodes associated with individuals, organizations, or computers and by
links representing their interactions. In many real networks, each link
has an associated weight, the larger the weight, the harder it is to
transverse the link. These networks are called ``weighted'' networks
\cite{Brauns03,Barat}.

Transport is one of the main functions of networks.  While the transport
on unweighted networks has been studied \cite{Lopez_transport}, the
effect of disorder on transport in networks is still an open question.
Here we study the distribution $P(\sigma)$ of the equivalent electrical
conductance $\sigma$ between two randomly selected nodes $A$ and $B$ on
Erd\H{o}s-R\'enyi (ER) \cite {E_R,E_R2}
and scale-free (SF) \cite {Albert02} weighted networks. We first
provide an iterative fast algorithm to obtain $P(\sigma)$ for disordered
resistor networks, and then we develop a theory to explain the behavior
of $P(\sigma)$.  The theory is based on the percolation theory
\cite{Bunde} for a weighted random network. We model a
weighted network by assigning the conductance of a link connecting node
$i$ and node $j$ as in Ref. \cite{Strelniker_2d_granular}
\begin{equation}\label{eq.sigma}
g_{ij}\equiv\exp[-ax_{ij}],
\end{equation}
where the parameter $a$ controls the broadness (``strength'') of the
disorder, and $x_{ij}$ is a random number taken from a uniform
distribution in the range [0,1]. We use this kind of disorder since a
recent study of magnetorresistance in real granular materials systems
\cite{Strelniker_2d_granular} shows that the conductance is given by
Eq.~(\ref{eq.sigma}).  Moreover, a recent study \cite{Chen_universal}
shows that many types of disorder distributions lead to the same
universal behavior.  The range of $a\gg 1$ is called the strong disorder
(SD) limit \cite{Cieplak, Porto}.  The special case of unweighted
networks, i.e., $a=0$ or $g_{ij}=1$ for all links have been studied earlier
\cite{Lopez_transport}.

To construct ER networks of size $N$, we randomly connect nodes with
$\av{k}N/2$ links, where $\av{k}$ is the average degree of the network.
To construct SF networks, in which the degree
distribution follows a power law, we employ the
Molloy-Reed algorithm \cite{Molloy}. The traditional algorithm to
calculate the probability density function (pdf) $P(\sigma)$ is to
compute $\sigma$ between two nodes $A$ and $B$ by solving the Kirchhoff
equations with fixed potential $V_A=1$ and $V_B =0$ and compute
$P(\sigma) d \sigma$, which gives the probability that two nodes in the
network have conductance between $\sigma$ and $\sigma + d \sigma$.
However, this method is time consuming and limited to relatively small
networks.  Here we also use an iteration algorithm proposed by
Grimmett and Kesten \cite{net math} to calculate $P(\sigma)$ and show
that it gives the same results as the traditional Kirchhoff method.

\begin{figure}[ht]
\includegraphics[width=0.4\textwidth]{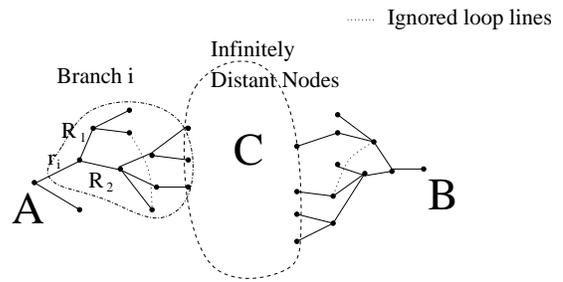}
\caption{Schematic Iteration model. In this example $R_1$ is infinite,
so it is not taken into account in the sum in $R_i$ of Eq.~(\ref{eq.branch R}).}
\label{it_graph}
\end{figure}

In the limit $N\to\infty$ we ignore the loops between 2 randomly
chosen nodes because the probability to have loops is very small. Hence
the resistivity $R_i$ of a randomly selected branch $i$ connecting a
node with infinitely distant nodes satisfies
$R_i=r_i+1/(\sum_{j=1}^{k-1}R_j^{-1})$, where $r_i=e^{ax_i}$ is the random
resistance of the link outgoing from this node
and $k$ is a random number taken from the distribution
$\tilde{p}_k=p_k\cdot k/\av{k}$, which is the probability that 
a randomly selected link ends in a node of
degree $k$, where $p_k$ is the original degree
distribution.  In Fig.~\ref{it_graph}, we show the schematic iteration
method.  The randomly selected nodes A and B are connected to the
infinitely distant nodes C. When we calculate $R_{AC}$, the resistance
between A and C, we perform the iterative steps as follows:

First we calculate the distribution of resistivities of the branches
connecting node A with C.  We start with
$\cal{N}$ branches having resistivities $R_i^{(0)}=0 \quad
(i=1,2,...,\cal{N})$, where $\cal{N}$ is an arbitrary large number.
Thus, initially the histogram of these resistivities $P_0(R)=\delta(R)$.
At the iterative step $n+1$, we compute a new histogram $P_{n+1}(R)$
knowing the histogram $P_n(R)$.  In order to do this we generate a new
set of resistivities $R_i^{(n+1)}$ by connecting in parallel $k-1$
outgoing branches coming from a randomly selected node of degree $k$
obtained from the distribution $\tilde{p}_k=p_k\cdot k/\av{k}$.  Then we
compute the resistivity of a branch going through this node via an
incoming link with a random resistivity $r_i^{(n)}$ taken from the link
resistivity distribution,
\begin{equation}\label{eq.branch R}
R_i^{(n+1)}=r_i^{(n)}+{1\over\sum_{j=1}^{k-1} 1/R_j^{(n)}}.
\end{equation}
In Eq.~(\ref{eq.branch R}), if at least one of the terms $R_i^{(n)}=0$,
we take $R_i^{(n+1)}=r_i^{(n)}$.  Thus after the first iterative step
$P_1(R)$ coincides with the distribution of link resistivities.

According to the theorem proved in \cite{net math}, as $n \to \infty$,
$P_n(R)$ converges to a distribution of the resistivities of a branch
connecting a node to the infinitely distant nodes.
The resistivity between a randomly selected node of degree $k$ and the
infinitely distant nodes is defined by
\begin{equation}
\tilde{R}_{(i)}={1\over\sum_{j=1}^k 1/R_j},
\end{equation}
where $k$ is selected from the original degree distribution $p_k$ and
$R_j$ is selected from $P_{n \to \infty}(R)$.

Finally, to compute the resistivity $R_{ij}$ between two randomly selected nodes
$i$ and $j$ (for example $A$ and $B$ in Fig.~\ref{it_graph}), we compute
$R_{ij}=\tilde{R}_{(i)}+\tilde{R}_{(j)}$,
where $\tilde{R}_{(i)}$ and $\tilde{R}_{(j)}$ are randomly selected
resistivities between a node and the infinitely distant nodes.
If $\cal{N}$ is a sufficiently large number, we find the conductance
distribution $P(\sigma)$ between any two randomly selected nodes.

\begin{figure}[ht]
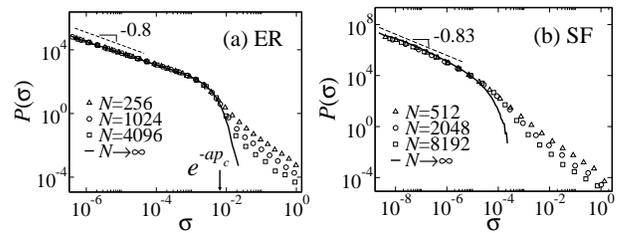

\includegraphics[width=0.22\textwidth]{diff_NER.eps}
\includegraphics[width=0.22\textwidth]{diff_NSF.eps}
\caption{Plots of $P(\sigma)$ for several values
of $N$. The symbols are for the Kirchhoff method and the solid line
is for the iterative method with $N \to \infty$.
(a) ER networks with fixed $\av{k}=3$ and
$a=15$. (b) SF networks with fixed $\lambda=3.5$, $k_{\rm min}=2$,
$\av{k} \approx 3.33$ and $a=20$.  The dashed line slopes are from the
prediction of Eq.(\ref{eq.P_sigma}) or (\ref{eq.p(sigma)}).
\label{fg_N}}
\end{figure}

In Figs.~\ref{fg_N}(a) and~\ref{fg_N}(b) we show the results of
$P(\sigma)$ using the traditional method of solving
Kirchhoff's equations for different values of $N$ and the iterative
method with $N\to\infty$ for both ER and SF networks. We see
that the main part of the distribution (low conductances) does not
depend on $N$, and only the high conductance tail depends on $N$.

The behavior of the two regimes, low conductance and high conductance,
can be understood qualitatively as follows: For strong disorder $a \gg 1$
all the current between two nodes follows the optimal path between them.
The problem of the optimal path in a random graph in the strong 
disorder limit can be mapped onto a percolation problem on a Cayley tree 
with a degree distribution identical to the random graph and with a fraction
$p$ of its edges conducting \cite{santafe}.
However, the conductance on this path is determined by the bond with
the lowest conductance $e^{-ax_{max}}$, 
where $x_{max}$ is the maximum random number along the path.
In the majority of cases $x_{max} > p_c$, where $p_c$ is the critical
percolation threshold of the network,
and only when the two nodes both belong to the incipient infinite 
percolation cluster (IIPC) \cite{Bunde}, $x_{max} < p_c$.
Since the size of the IIPC scales as $N^{2/3}$,
the probability of randomly
selecting a node inside the IIPC is proportional to $N^{2/3}/N=N^{-1/3}$
\cite{E_R,E_R2,Bunde}. Then the probability of randomly selecting a pair inside the
IIPC is proportional to $(N^{-1/3})^2=N^{-2/3}$.
These nodes contribute to 
the high conductance range $\sigma > e^{-ap_c}$ of $P(\sigma)$.
The low conductance regime is determined by the
distribution of $x_{max}$, 
that follows the behavior of the order
parameter $P_{\infty}(p)$ (for $p>p_c$) in the percolation problem
which is independent of $N$ \cite{santafe}.
(This will be explained later in the theoretical approach for the low
conductance regime.)

We call the low conductance regime a {\it
non-percolation regime\/} and the high conductance regime a {\it
percolation regime}. In contrast, the property of existing two regimes
does not show up
in the optimal path length \cite{Tomer-op,Wu_flow} and only the scaling
regime with $N$ appears. This is since the path length for almost all
pairs is dominated by the IIPC \cite{Wu_flow}.

\begin{figure}[ht]
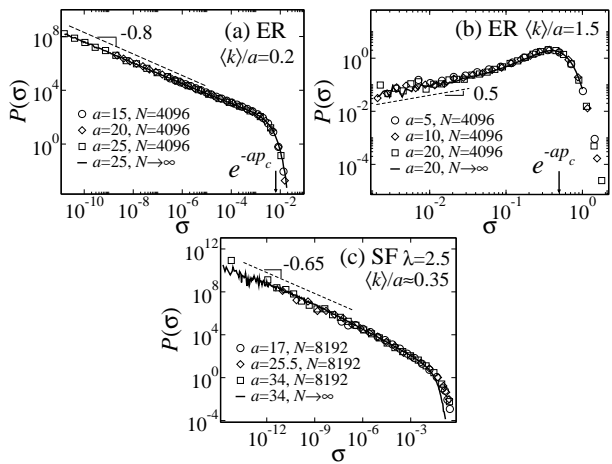

\includegraphics[width=0.22\textwidth]{er_ka0.2.eps}
\includegraphics[width=0.22\textwidth]{er_ka1.5.eps}
\includegraphics[width=0.22\textwidth]{sf_lam2.5_ka0.12.eps}
\caption{Plots of $P(\sigma)$ for fixed $\av{k}/a$.
The symbols are for the Kirchhoff method and the solid line
is for the iterative method.
For the same $\av{k}/a$, the iterative method for
different $a$ shows the same $P(\sigma)$ except that the lower cutoff
is different.
(a) ER network with $\av{k}/a=0.2$.
(b) ER network with $\av{k}/a=1.5$.
(c) SF network with $\av{k}/a\approx 0.35$, $\lambda =2.5$.
The dashed line slopes are from the
prediction of Eq.(\ref{eq.P_sigma}) or (\ref{eq.p(sigma)}).
\label{fg_ka}}
\end{figure}

In Figs.~\ref{fg_ka}(a) and \ref{fg_ka}(b) we plot for a given $N$ only
the non-percolation part of $P(\sigma)$ as a function of $\sigma$ for
fixed values of $\av{k}/a$ and different $\av{k}$ and $a$ values for
ER networks. We see that it obeys a power law with the slope
$\av{k}/a-1$ for $\sigma < e^{-a p_c}$.
Note that for ER networks $p_c=1/\av{k}$ \cite{E_R,E_R2}. In
Fig.~\ref{fg_ka}(c), we plot the conductance distribution for SF
networks for fixed values of $\av{k}/a$.  We can see the non-percolation part
seems to obey the same power law as ER networks.

Next we present an analytical approach for the form of $P(\sigma)$
for low conductance regime. The distribution of the maximal
random number $x_{\rm max}$ along the optimal path can be expressed in
terms of the order parameter $P_{\infty}(p)$ in the percolation problem
on the Cayley tree, where $P_{\infty}(p)$
is the probability that a randomly chosen node on the Cayley tree 
belongs to the IIPC \cite{santafe}.
For a random graph with degree distribution $p_k$, the probability
to arrive at a node with $k$ outgoing branches by following a randomly
chosen link is $(k+1)p_k/\av{k}$ \cite{branch theo}.  The probability
that starting at a randomly chosen {\bf link} on a Cayley tree one can reach
the $\ell$th generation is
\begin{equation}\label{eq.fl}
f_{\ell}(p) \equiv f_{\ell}=1-\sum_{k=1}^\infty \frac{p_k k \left(1- p f_{{\ell}-1}\right)^{k-1}}{\langle k \rangle},
\end{equation}
where $f_0=1$. Slightly different from $f_{\ell}$ is the probability
that starting at a randomly chosen {\bf node} one can reach the $n$th
generation,
\begin{equation}\label{eq.fl_tilde}
\tilde{f}_n=1-\sum_{k=0}^\infty p_k (1- p f_{n-1})^k.
\end{equation}
In the asymptotic limit $f_{\ell}$ converges to $P_\infty$ for a
given value of $p$,
\begin{equation}\label{eq.pinf}
f_{\ell}\to P_\infty(p)=1-\sum_{k=1}^\infty \frac{p_k k \left(1- p P_\infty
\right)^{k-1}}{\langle k \rangle}.
\end{equation}
In this limit we have a pair of nodes on a random graph separated by a
very long path of length $n$.  The probability that two nodes will be
connected (conducting) at given $p$, can be approximated by the
probability that both of them belong to the IIPC
\cite{net math}:
\begin{equation}\label{eq.pip}
\Pi(p) = \left[{\tilde{P}_\infty(p)\over\tilde{P}_\infty(1)}\right]^2,
\end{equation}
where $\tilde{P}_\infty(p)\equiv \lim_{n \to \infty}
\tilde{f}_n=1-\sum_{k=0}^\infty p_k (1- p P_\infty)^k$.  Note that the
negative derivative of $\Pi(p)$ with respect to $p$ is the distribution
of $x_{max}$ and thus gives $P(\sigma)$ in the SD limit.
In our case $\sigma=e^{- ap}$,
 so replacing $p$ by $p=-\ln \sigma/a$ in Eq.~(\ref{eq.pip}) and
differentiating with respect to $\sigma$, we obtain the distribution of
conductance in the SD limit when the source and sink are far apart ($n
\to \infty$),
\begin{equation}\label{eq.psigma}
P(\sigma)=-{d\over d\sigma}\Pi(\sigma)=\frac{2 \tilde{P}_\infty (p)}{\sigma a
[\tilde{P}_\infty (1)]^2}\cdot \frac{\partial \tilde{P}_\infty(p)}{\partial p}~\mid_{p=-\ln\sigma/a}.
\end{equation}

For ER networks the degree distribution is a Poisson distribution
with $p_k=\langle k \rangle ^k e^{- \langle k \rangle}/k!$ \cite{E_R,E_R2}
and thus $P_\infty(p)$ satisfies
\begin{equation}\label{eq.ER_P}
P_\infty(p)=1-e^{- \langle k \rangle p P_\infty(p)},
\end{equation}
which has a positive root $P_\infty$ for $p>p_c=1/\av{k}$.
And $\tilde{P}_\infty (p) = P_\infty (p)$, thus
\begin{equation}\label{eq.pi}
P(\sigma)=\frac{2 P_\infty (p)}{\sigma a [P_\infty (1)]^2}\cdot \frac{\partial P_\infty(p)}{\partial p}~\mid_{p=-\ln\sigma/a},
\end{equation}
where $P_\infty(p)$ and $P_\infty(1)$ are the solutions of
Eq.~(\ref{eq.ER_P}).

We test the analytical result Eq.~(\ref{eq.pi}) by comparing the
numerical solution of Eqs.~(\ref{eq.ER_P}) and~(\ref{eq.pi}) with the
simulations on actual random graphs by solving Kirchhoff equations
(Figs.~\ref{fg_N} and~\ref{fg_ka}).
The agreement between the simulations
and the theoretical prediction is perfect in the SD limit, i.e. when
$\av{k}/a$ is small.

\begin{figure}[ht]
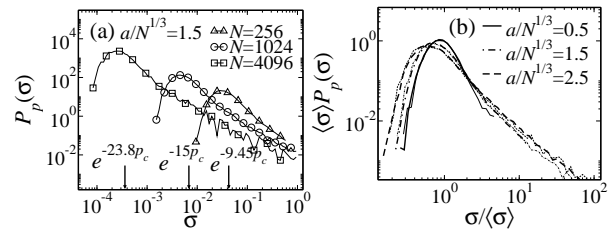

\includegraphics[width=0.22\textwidth]{k3_perc.eps}
\includegraphics[width=0.22\textwidth]{k3_perc_scaled.eps}
\caption{Kirchhoff method results of the percolation part of ER networks
with the same value of $p_c=1/\av{k}=0.33$.
(a) Normalized $P_p(\sigma)$ for fixed $a/N^{1/3}=1.5$.  (b) Scaled plot
of $\av{\sigma}P(\sigma)$ as function of $\sigma/\av{\sigma}$ for three
values of $a/N^{1/3}$. For each value of $a/N^{1/3}$, the thick line is for
$N=256$ and the thin line is for $N=1024$.
\label{ER_perc}}
\end{figure}

Next we simplify $P(\sigma)$ from Eq.~(\ref{eq.pi}).  Assuming
that $P_\infty(1) \approx 1$ which is true for large $\av{k}$
and approximating a slow varying function $P_{\infty}(p)$ by $P_{\infty}(1)$
we obtain
\begin{equation}\label{eq.P_sigma}
P(\sigma) \approx 2 \frac {\av{k}}{a} \sigma^{\av{k}/a-1},
\end{equation}
for the range $e^{-a} \le \sigma \ll e^{-ap_c}$ with $p_c=1/\av{k}$.  In
Figs.~\ref{fg_N} and~\ref{fg_ka} we also show the results predicted by
Eq.~(\ref{eq.P_sigma}).  For an infinite network, for $p \le
p_c=1/\av{k}$, $P_\infty(p)=0$, and hence, the distribution of
conductances must have a cutoff at $\sigma=e^{-ap_c}$.  Indeed, in
Fig.~\ref{fg_N}(a) and Figs.~\ref{fg_ka}(a) and \ref{fg_ka}(b) we see that
the upper cutoff of the iterative curves is close to $e^{-ap_c}$.

As discussed above, the range of high conductivities
corresponds to the case where both the
source and the sink are on the IIPC.  Previously we found this
percolation part to scale as $N^{-2/3}$.  Using Fig.~\ref{fg_N}(a), we
compute the integral for each $P(\sigma)$ from $e^{-ap_c}$ to $\infty$,
and find that indeed $\int_{e^{-ap_c}}^{\infty} P(\sigma)\mathrm{d}
\sigma \sim N^{-2/3}$, in good agreement with the theoretical approach.
To show how the percolation part of $P(\sigma)$ is related to the
parameters $N$, $a$ and $p_c$, we analyze the conductance between pairs on
the IIPC, i.e., each link on the optimal path from source to sink
has $x$ less than $p_c$. We compute $P_p(\sigma)$ of these
pairs on the IIPC. When we simulate this process, we have
only $N^{-2/3}$ probability to find this part from the original
normalized distribution $P(\sigma)$.  Thus, we normalize $P_p(\sigma)$
by dividing by $N^{-2/3}$.  Figures~\ref{ER_perc}(a) and
\ref{ER_perc}(b) show the normalized $P_p(\sigma)$ of pairs on the IIPC. 
In this range, we see that $P_p(\sigma)$ is
dominated by high conductivities and we find $\av{\sigma} \approx
e^{-ap_c}$ and
\begin{equation}
\label{e.1x}
\av{\sigma}P_p(\sigma)=f\left({\sigma\over\av{\sigma}},{ap_c\over
N^{1/3}}\right),
\end{equation}
that is, for fixed $ap_c/N^{1/3}$, $\av{\sigma}P_p(\sigma)$ scales with
$\sigma/\av{\sigma}$ as seen in Fig.~\ref{ER_perc}(b). The scaled
distributions have the same shape for the same $ap_c/N^{1/3}$ which
specifies the strength of disorder similarly to the behavior of the
optimal path lengths \cite{Chen_universal,Tomer-op,Wu_flow,sameet_opt}.
The explanation of this fact for the distribution of conductances is
analogous to the arguments presented in Refs. \cite{santafe} and
\cite{Tomer-op} for the distribution of the optimal path.
Thus the position of the
maximum of the scaled curves in Fig.~\ref{ER_perc}(b), and the whole
shape of the distributions, depend on $ap_c/N^{1/3}$.

We also find that the extreme high conductivities correspond to the case
where source and sinks are separated by only one link.
In this case, $P(\sigma)=\frac{\av{k}}{aN\sigma} \sim \sigma^{-1}$,
($\sigma<1$).

In summary, we find that $P(\sigma)$ exhibits two regimes. For $\sigma<e^{-ap_c}$,
we show both analytically and numerically that for ER networks
$P(\sigma)$ follows a power law,
\begin{equation}\label{eq.p(sigma)}
P(\sigma)\sim \sigma^{-\alpha}\qquad [\alpha=1-\av{k}/a].
\end{equation}
We also find that for SF networks, Eq.~(\ref{eq.p(sigma)}) seems to
be a good approximation, consistent with numerical simulations.
The distributions of optimal path length and the path length of the
electrical currents in complex weighted networks \cite{Tomer-op,Wu_flow}
have been found to depend on
$N$ for all length scales and all types of networks studied.  In
contrast, here we find that the {\it low\/} conductance tail of
$P(\sigma)$ does not depend on $N$ for both ER and SF networks.
However, the {\it high\/} conductance regime ($\sigma > e^{-ap_c}$) of
$P(\sigma)$ does depend on $N$, in a similar way to the optimal path
length and current path length distributions \cite{Tomer-op,Wu_flow}.

We thank ONR, Dysonet, FONCyt (PICT-O 2004/370), FONCyt (PICT-O 2004/370),
Israel Science Foundation and Conycit for support, and Zhenhua Wu for
helpful discussions.


\begin{thebibliography}{99}

\bibitem{Albert02} R. Albert and A.-L. Barab\'{a}si,
Rev. Mod. Phys. {\bf 74}, 47 (2002).

\bibitem{Dorogovtsev} S. N. Dorogovtsev and J. F. F. Mendes,
\textit{Evolution of Networks: From Biological Nets
to the Internet and WWW} (Oxford University Press, Oxford, 2003).

\bibitem{Pastor} R. Pastor-Satorras and A. Vespignani, \textit{Structure
and Evolution of the Internet: A Statistical Physics Approach}
(Cambridge University Press, Cambridge, 2004).

\bibitem{Cohen} R. Cohen and S. Havlin, \textit{Complex networks:
Stability, Structure and Function} (Cambridge University Press,
Cambridge, In press).

\bibitem{Brauns03} L. A. Braunstein {\it et al.},
Phys. Rev. Lett. {\bf 91}, 168701 (2003).

\bibitem{Barat} A. Barrat, M. Barth\'elemy, R. Pastor-Satorras and
A. Vespignani, PNAS {\bf 101}, 3747
(2004).

\bibitem{Lopez_transport} E. L\'opez {\it et al.}, Phys. Rev. Lett. {\bf
94}, 248701 (2005).

\bibitem{E_R} P. Erd\H{o}s and A. R\'enyi, Publ. Math. (Debrecen) {\bf
6}, 290 (1959).

\bibitem{E_R2}  P. Erd\H{o}s and A. R\'enyi, Publications of the Mathematical Inst.
of the Hungarian Acad. of Sciences {\bf 5}, 17 (1960).


\bibitem{Bunde} A. Bunde and S. Havlin, \textit{Fractals and Disordered
Systems\/} (Springer-Verlag, Heidelberg, 1995).

\bibitem{Strelniker_2d_granular} Y. M. Strelniker {\it et al.},
Phys. Rev. E {\bf 69}, 065105(R) (2004).

\bibitem{Chen_universal} Y. Chen {\it et al.}, Phys. Rev. Lett. {\bf 96}, 068702 (2006).

\bibitem{Cieplak} M. Cieplak {\it et al.}, Phys. Rev. Lett. {\bf 72}, 2320 (1994); {\bf 76}, 3754 (1996).

\bibitem{Porto} M. Porto {\it et al.},  Phys. Rev. E {\bf 60}, R2448 (1999).

\bibitem{Molloy} M. Molloy and B. Reed, Random Structures and Algorithms
{\bf 6}, 161 (1995); Combin. Probab. Comput. {\bf 7}, 295 (1998).

\bibitem{net math} G. Grimmett and H. Kesten,
 \textit{Random electrical networks on complete graphs II: Proofs}, 1983
(http://arxiv.org/abs/math.PR/0107068).

\bibitem{santafe} L. A. Braunstein {\it et al.}, in {\it Lecture
  Notes in Physics: Proceedings of the 23rd CNLS Conference, ``Complex
    Networks,'' Santa Fe 2003}, edited by E. Ben-Naim, H. Frauenfelder,
    and Z. Toroczkai (Springer, Berlin, 2004).

\bibitem{Tomer-op} T. Kalisky {\it et al.}, Phys. Rev. E {\bf 72}, 025102(R) (2005).

\bibitem{Wu_flow} Z. Wu {\it et al.}, Phys. Rev. E {\bf 71},
045101(R) (2005).

\bibitem{branch theo} T. E. Harris, \textit{The Theory of Branching Processes\/}
(Dover Publication Inc., New York, 1989).

\bibitem{sameet_opt} S. Sreenivasan {\it et al.}, Phys. Rev. E {\bf 70},
046133 (2004).



\end{thebibliography}
\end{document}